\documentclass[reprint,showpacs,preprintnumbers,amsmath,amssymb,pre]{revtex4-1}
\usepackage{graphicx,wasysym}
\usepackage{bm}
\usepackage{tabulary}

\newcommand{\kB}{k_\mathrm{B}}

\newcommand{\Nw}{N_{\mathrm{w}}}

\makeatletter
\newcommand*{\inlineequation}[2][]{%
  \begingroup
    \refstepcounter{equation}%
    \ifx\\#1\\%
    \else
      \label{#1}%
    \fi
    \relpenalty=10000 %
    \binoppenalty=10000 %
    \ensuremath{%
      #2%
    }%
    ~\@eqnnum
  \endgroup
}
\makeatother

\begin{document}

\preprint{1}

\title{Transport Coefficients from Large Deviation Functions}

\author{Chloe Ya Gao$^1$}
\author{David T. Limmer$^{1,2,3}$} \email{dlimmer@berkeley.edu}

\affiliation{%
$^{1)}$ Department of Chemistry, University of California, Berkeley, CA 94609 \\
$^{2)}$ Kavli Energy NanoScience Institute, Berkeley, CA 94609\\
$^{3)}$ Materials Science Division, Lawrence Berkeley National Laboratory, Berkeley, CA 94609
}

\date{\today}
\begin{abstract}
We describe a method for computing transport coefficients from the direct evaluation of large deviation function. This method is general, relying on only equilibrium fluctuations, and is statistically efficient, employing trajectory based importance sampling. Equilibrium fluctuations of molecular currents are characterized by their large deviation functions, which is a scaled cumulant generating function analogous to the free energy.  A diffusion Monte Carlo algorithm is used to evaluate the large deviation functions, from which arbitrary transport coefficients are derivable. We find significant statistical improvement over traditional Green-Kubo based calculations. The systematic and statistical errors of this method are analyzed in the context of specific transport coefficient calculations, including the shear viscosity, interfacial friction coefficient, and thermal conductivity.
\end{abstract}

\maketitle


The evaluation of transport coefficients from molecular dynamics simulations is a standard practice throughout physics and chemistry. 
Despite significant interest and much study, such calculations remain computationally demanding. 
Traditional methods exploit Green-Kubo relationships \cite{green1954markoff,kubo1957statistical1} and rely on integrating equilibrium time correlation functions \cite{levesque1973computer,schelling2002comparison,galamba2004thermal,jones2012adaptive}. 
While general, depending only on identifying a relevant molecular current, these methods often suffer from large statistical errors due to finite time averaging, making them cumbersome to converge \cite{evans1978transport,hess2002determining}. 
Alternative methods exist that directly drive a current through the system by the application of specific boundary conditions \cite{tenenbaum1982stationary,baranyai1999steady} or by altering the equations of motion \cite{hoover1980lennard,evans1982homogeneous,mandadapu2009homogeneous,muller1997simple}. 
These direct methods typically mitigate sampling difficulties by requiring that only the current is averaged rather than its time correlation function. 
However, such methods are generally not transferable to different transport processes. Moreover, as a nonequilibrium simulation, the details of how the current is generated can affect their convergence \cite{zhou2009towards} and fidelity \cite{tuckerman1997modified,tenney2010limitations}. 
Here we propose a new way to compute transport coefficients that utilizes only equilibrium fluctuations, as in Green-Kubo calculations, but is evaluated by averaging a current, as in direct methods. Rather than applying a physical field to drive a current, we apply a statistical bias to the system's dynamics and measure the resultant response. 
The response is codified in the relative probability of a given current fluctuation, so this calculation is identical to the evaluation of a free energy, albeit in a path ensemble \cite{geissler2004equilibrium}. Such path ensemble free energies are known as large deviation functions \cite{touchette2009large}, and with trajectory based importance sampling methods to aid in their calculation, we arrive at a method to evaluate transport coefficients that is both general and quickly convergent. 

Large deviation theory has emerged as a useful formalism for considering the fluctuations of time integrated observables \cite{touchette2017introduction}. 
In fact, large deviation theory underpins many recent developments in nonequilibrium statistical mechanics, including generalized fluctuation theorems \cite{jarzynski1997nonequilibrium,crooks1999entropy} and thermodynamic uncertainty principles \cite{barato2015thermodynamic,gingrich2016dissipation}. 
The large deviation function is a scaled cumulant generating function and, like its equilibrium counterpart, the free energy, it codifies the stability and response of dynamical systems. 
While these advances in nonequilibrium statistical mechanics have yielded important relationships for systems far from equilibrium, they have also brought new insight into near-equilibrium phenomena. 
Andrieux and Gaspard have illustrated this especially clearly, resolving how Onsager's reciprocal relations and their generalizations beyond linear response follow from the large deviation function for the total entropy production and its symmetry provided by the fluctuation theorem \cite{gaspard2013multivariate}.
They have shown the connection between the moments of a large deviation function for a time integrated current and phenomenological transport coefficients within both linear and nonlinear response regimes \cite{andrieux2004fluctuation,andrieux2007fluctuation}. 
We use this insight--that large deviation functions can encode the dynamical response of a system driven away from equilibrium--to construct an efficient method for the evaluation of transport coefficients from molecular dynamics simulations. 

To compute a large deviation function for a time integrated molecular current, we employ a trajectory based importance sampling procedure. 
Beginning with transition path sampling \cite{dellago1998transition}, Monte Carlo algorithms have been derived to uniformly sample path space for systems evolving with detailed balanced dynamics. 
These methods have found application in computing rate constants and finding reaction pathways for complex condensed phase processes spanning autoionization to viral capsid assembly \cite{geissler1999kinetic,geissler2001autoionization,bolhuis2002transition,radhakrishnan2004orchestration,basner2005enzyme,hagan2006dynamic,peters2010recent,limmer2014theory}. 
Indeed, it was identified in early work that a reaction rate constant could be computed from a thermodynamic-like integration along path space, resulting in a relative free energy in trajectory space  \cite{dellago1998transition,geissler2004equilibrium}.
This observation was never generalized to other dynamical responses, like phenomenological transport coefficients.
With the development of diffusion Monte Carlo algorithms like the cloning algorithm that directly target large deviation functions \cite{giardina2006direct,giardina2011simulating}, such generalization is possible. 
Recent extensions of diffusion Monte Carlo algorithms that incorporate importance sampling using an iterative feedback approach \cite{nemoto2016population}, cumulant expansions \cite{klymko2017rare}, or approximate auxiliary processes \cite{ray2017exact}, have improved the efficiency of these algorithms enough to make calculations for complex, high dimensional systems possible. 
In this way, we proceed numerically by computing directly an effective thermodynamic potential like that Onsager identified when he first formulated his thermodynamic theory of linear response \cite{onsager1931reciprocal}, with the large deviation function serving to characterize this potential. 
This conceptually distinct approach from traditional methodologies provides new ways of thinking about transport processes that are amenable to the kinds of analysis typically reserved for static equilibrium observables, such as their dependence on ensemble and generalization to nonlinear regimes\cite{palmer2017thermodynamic}. While we are restricted to linear response coefficients in this article, generalization to nonlinear response regimes is straightforward \cite{andrieux2007fluctuation}.

The rest of the paper is organized in the following manner. In Section 2, we summarize important results of the large deviation theory, and illustrate its connection to phenomenological transport coefficients. We also outline the simulation methodology used to compute large deviation functions. In Section 3, we test our method by comparing it with calculations using the Green-Kubo formalism. We study three specific cases: the shear viscosity of TIP4P/2005 water \cite{abascal2005general}, the interfacial friction coefficient between a Lennard-Jones fluid and a Lennard-Jones wall, and the thermal conductivity of a Weeks-Chandler-Anderson \cite{weeks1971role} solid. We use these models to frame a discussion of the relative systematic and statistical errors associated with our new methodology in comparison to traditional Green-Kubo calculations. We provide some final remarks on our method in Section 4.

\section{Theory and Methodology}

We consider systems evolving according to a Markovian stochastic dynamics, though generalization to deterministic dynamics is straightforward. In the absence of an external stimulus, these dynamics obey microscopic reversibility and thus will sample a Boltzmann distribution \cite{chandler1987introduction}. Under a bias, applied either at the boundaries of the system or through an external field, a current is expected to arise. If the bias is small, the response of the system can be linearized and a transport coefficient, $L$, is  defined through
\begin{equation}
\label{Eq:JLX}
J = LX \, ,
\end{equation}
where $J$ is a current and $X$ is its conjugate generalized force, which could be proportional to a temperature or concentration gradient. The entropy production for this process is equal to the product of the force and the current, or $S = J X$ \cite{morriss2013statistical}. The transport coefficient, $L$, is thus a response function relating the applied force to the generated current, $L=dJ/dX$, in the limit that $X \rightarrow 0$. This is the object we aim to compute.

\subsection{Transport coefficients from large deviation functions}

To compute the response coefficient, $L$, we must identify a corresponding dynamic variable whose fluctuations will report on the system's response to the bias. Specifically, we define a time averaged current as  
\begin{equation}
J=\frac{1}{t_N}\int_{0}^{t_N} j(c_t) dt,
\end{equation}
where $t_N$ is some observation time, and $j(c_t)$ is a fluctuating variable computable from the molecular configuration, $c_t$, at time $t$. If $j$ is correlated over a finite amount of time, then the fluctuations of $J$ can be studied by computing its cumulant generating function,

\begin{equation}
\label{Eq:LDF}
\psi (\lambda) = \lim_{t_N\rightarrow \infty }\frac{1}{t_N} \ln\left \langle e^{-\lambda t_NJ} \right \rangle.
\end{equation}
where $\psi (\lambda)$ is known as the large deviation function and $\lambda$ is a statistical field conjugate to $J$ \cite{touchette2009large}. Here, the average $\left \langle \cdots \right \rangle$ is taken within an ensemble of paths of length $t_N$, denoted as a vector of all the configurations visited over that time, or $C(t_N)=\left \{ c_0,c_1,\cdots ,c_{t_N} \right \}$. The probability of observing such a path is given by, 
\begin{equation}
P[C(t_N)]=\rho [c_0] \prod_{i=1}^{t_N}\omega[c_{i-1}\rightarrow c_i],
\end{equation}
where $\rho [c_0]$ represents the distribution of initial conditions, and $\omega[\cdots]$ are the transition probabilities between time-adjacent configurations. 

As a cumulant generating function, the derivatives of $\psi (\lambda )$ report on the fluctuations of the current $J$. For example, the first two derivatives yield
\begin{eqnarray}
\label{Eq:Clts}
\psi'(0) &= -\left \langle J \right \rangle, \nonumber \\
\psi''(0) &= t_N\left \langle \delta J^2 \right \rangle,
\end{eqnarray}
where  $\left \langle J \right \rangle$ is the average current and $\delta J = J - \left \langle J \right \rangle$ is its deviation from the mean, whose squared average yields the variance of $J$. Because the dynamics we consider obey microscopic reversibility, $\psi (\lambda )$ obeys a generalized fluctuation theorem, $\psi (\lambda )=\psi (X-\lambda )$ where $X$ is the generalized force as in Eq.~\ref{Eq:JLX}. This symmetry relates the likelihood of a current to its time-reversed conjugate \cite{andrieux2004fluctuation,lebowitz1999gallavotti}, and implies a fluctuation-dissipation relationship, or a relation between the second derivative of the large deviation function
\begin{equation}
 \psi^{''}(\lambda) = -\frac{\partial \left \langle J \right \rangle_\lambda }{\partial \lambda} = 2 \frac{\partial \left \langle J \right \rangle _\lambda}{\partial X} \, ,
\end{equation}
and the transport coefficient, $L$. Here $\langle \dots \rangle_\lambda$ denotes the average in the biased path ensemble. In the limit of $X \rightarrow 0$,
\begin{equation}
\psi^{''}(0) =2 L\, ,
\end{equation}
where as previously observed \cite{andrieux2004fluctuation}, we find that the curvature of the large deviation function around $\lambda=0$ is equal to the response function $L$ up to a factor of 2. Analogously, higher order derivatives can be related to nonlinear transport coefficients. For small values of $\lambda$, the large deviation function can be expanded as
\begin{equation}
\psi (\lambda) = L\lambda ^2 + O(\lambda ^4),
\end{equation}
which is parabolic, and completely determined by $L$. This implies the distribution of $J$ is Gaussian, with a variance of $2L/t_N$. This inversion is a direct reflection of Onsager's notion of an effective thermodynamic potential, where the probability of a current is given by the exponential of the entropy production.

The connection between the large deviation result and the Green-Kubo formalism can be understood by expanding the definition of $J$ in Eq. ~\ref{Eq:Clts}. Without loss of generality, within Green-Kubo theory a transport coefficient can be computed from,
\begin{equation}
\label{Eq:Tm}
L = \lim_{t_M \rightarrow \infty} L(t_M),\quad \quad L(t_M)=\int_{0}^{t_M} \langle j(c_0)j(c_t)  \rangle dt \, ,
\end{equation}
where $L(t_m)$ is an integral over the time correlation function of $j(c_t)$, and in the long time limit is equal to $L$ \cite{morriss2013statistical}. As $\left \langle J \right \rangle=0$ for an equilibrium system, where $X=0$, it is straightforward to relate the second derivative of the large deviation function with respect to $\lambda$ evaluated at $\lambda=0$, to $L$ as
\begin{equation}
\label{Eq:EH}
\psi^{''}(0) = 2\int_{0}^{\infty}\left \langle j(c_0)j(c_t) \right \rangle dt = 2 L \, ,
\end{equation}
where we have made use of the time-translational invariance of the equilibrium averaged time correlation function, and assumed that $\left \langle j(c_0)j(c_t) \right \rangle$ decays faster than $1/t$. This equation is known as the Einstein-Helfand relation and is well known to yield an equivalent expression for transport coefficients \cite{helfand1960transport}. Provided an estimate of $\psi(\lambda)$ accurate enough to compute $\psi^{''}(0)$, we thus have a means of evaluating $L$.

\subsection{Calculation of large deviation functions}

To evaluate the large deviation function for $J$, we use a variant of path sampling known as the cloning algorithm \cite{giardina2006direct,giardina2011simulating}. The cloning algorithm is based on a diffusion Monte Carlo procedure \cite{foulkes2001quantum} where an ensemble of trajectories is integrated in parallel. Each individual trajectory is known as a walker, and collectively the walkers undergo a population dynamics whereby short trajectory segments are augmented with a branching process that results in walkers being pruned or duplicated in proportion to a weight. This algorithm has been used extensively in the study of driven lattice gases \cite{hurtado2011spontaneous} and models of glasses \cite{garrahan2009first,bodineau2012finite}. Alternative methods for importance sampling trajectories, such as transition path sampling \cite{bolhuis2002transition} or forward flux sampling \cite{allen2009forward}, could be used similarly.

Generally, to importance sample large deviation functions, the original trajectory ensemble, $P[C(t_N)]$, can be biased to the form \cite{frenkel2001understanding}
\begin{equation}\label{Eq:Plambda}
P_\lambda[C(t_N)] = P[C(t_N)]e^{-\lambda t_N J[C(t_N)]-\psi(\lambda) t_N} \, ,
\end{equation}
where the large deviation function $\psi(\lambda)$ is the normalization constant computable as in Eq.~\ref{Eq:LDF}. Ensemble averages for an arbitrary observable, $\mathcal{O}$, within the unbiased distribution and the biased one, are related by
\begin{equation}
\label{Eq:BiasedEn}
\langle  \mathcal{O}[C(t_N)] \rangle_\lambda = \frac{ \langle \mathcal{O}[C(t_N)]  e^{-\lambda t_N J[C(t_N)]} \rangle}{\langle e^{-\lambda t_N J[C(t_N)]} \rangle} \, ,
\end{equation}
where the denominator is $\exp[\psi(\lambda) t_N]$ in the limit of large $t_N$. If we choose $\mathcal{O}[C(t_N)] = \delta(J-J[C(t_N)])$ in Eq.~\ref{Eq:BiasedEn}, then we find a familiar relationship between biased ensembles,
\begin{equation}
\label{Eq:Rew}
\ln p_\lambda (J) = \ln p(J)-\lambda t_NJ-t_N \psi(\lambda).
\end{equation}
where $p_\lambda(J) =\langle \delta(J-J[C(t_N)]) \rangle_\lambda$ is the probability of observing a given value of the current $J$ in the biased ensemble, and $p(J)$ is that in the unbiased ensemble. This demonstrates that $\psi(\lambda)$ is computable as a change in normalization through histogram reweighting \cite{frenkel2001understanding}. 

\begin{table*}
\caption{Transport coefficients with corresponding Green-Kubo formula and dynamical observable.\vspace{0.2cm}}
\centering
\setlength\extrarowheight{10pt}
\begin{tabular}[t]{cll} \hline
\vspace{0.2cm}
\textbf{Transport coefficient}\,\,	 & \hspace{0.8cm}\textbf{Green-Kubo relation}	& \hspace{0.5cm}\textbf{Dynamical observable}\\ \hline
\vspace{0.2cm}
shear viscosity		& \(\displaystyle \inlineequation[eta] {\eta = \frac{V}{k_{\mathrm{B}}T }\int_{0}^{\infty}\left \langle \sigma _{xy}(0)\sigma_{xy}(t) \right \rangle dt\hspace{0.5cm}} \)	\,\,\,\,\,		& \(\displaystyle \inlineequation[Eta]{\Sigma _{xy}=\frac{1}{t_N}\int_{0}^{t_N}\sigma _{xy}(t)dt\hspace{0.5cm}} \) \\
\vspace{0.2cm}
interfacial friction coefficient	\,\,\,\,\,\,\,\,	& \(\displaystyle \inlineequation[mu]{\mu = \frac{A}{k_{\mathrm{B}}T }\int_{0}^{\infty}\left \langle f _x(0)f_x(t) \right \rangle dt\hspace{0.8cm}} \)	\,\,\,\,\,\,\,\,& \(\displaystyle \inlineequation[Mu]{F_x=\frac{1}{t_N}\int_{0}^{t_N}f_x(t)dt\hspace{0.85cm}} \)\\ 
\vspace{0.2cm}
thermal conductivity		& \(\displaystyle \inlineequation[kappa]{\kappa =\frac{1}{Vk_{\mathrm{B}}T ^2}\int_{0}^{\infty}\left \langle q_x(0)q_x(t) \right \rangle dt\hspace{0.45cm}} \)			& \(\displaystyle \inlineequation[Kappa]{Q_x=\frac{1}{t_N}\int_{0}^{t_N}q_x(t)dt\hspace{0.8cm}}\) \\ \hline
\vspace{0.2cm}
\end{tabular}
\end{table*}

In order to arrive at a robust estimate for $\psi(\lambda)$, the two distributions, $p_\lambda (J)$ and $p(J)$, must have significant overlap. However, for large systems or long observation times, each distribution narrows, and sampling $p_\lambda(J)$ by brute force is exponentially difficult. To evaluate the large deviation function, the cloning algorithm samples $P_\lambda[C(t_N)]$ by noting that it can be expanded to 
\begin{equation}\label{Eq:Plambda2}
P_\lambda[C(t_N)] \propto \rho [c_0] \prod_{i=1}^{t_N}\omega[c_{i-1}\rightarrow c_i] e^{-\lambda \delta t j[c_i] } \, ,
\end{equation}
where we have discretized the integral for $J$ over a time $\delta t$. The argument of the product is the transition probability times a bias factor that is local in time. This combination of terms cannot be lumped together into a physical dynamics, as it is unnormalized. However, it can be interpreted as a population dynamics where the nonconservative part proportional to the bias is represented by adding and removing walkers. In particular, in the cloning algorithm, trajectories are propagated in two steps. First, $\Nw$ walkers are integrated according to the normalized dynamics specified by $\omega[c_{i-1}\rightarrow c_i]$ for a trajectory of length $n \delta t$. Over this time, a bias is accumulated according to  
\begin{equation}
W_i(t,n \delta t) = \exp \left [ - \lambda \delta t \sum_{j=1}^n j[c_{t+j \delta t}] \right ],
\end{equation}
where, due to the multiplicative structure of the Markov chain, the bias is simply summed in the exponential. After the trajectory integration, $n_i(t)$ identical copies of the $i$th trajectory are generated in proportion to $W_i(t,n \delta t)$,
\begin{equation}
\label{Eq:weights2}
n_i(t) = \left \lfloor N_\mathrm{w} \frac{W_i(t,n \delta t) }{\sum_{j=1}^{N_\mathrm{w}} W_j(t,n \delta t)}  + \xi \right \rfloor,
\end{equation}
where $\xi$ is a uniform random number between 0 and 1 and $\lfloor \dots \rfloor$ is the floor function. This process will result in a different number of walkers, and thus each walker in the new population is copied or deleted uniformly until $\Nw$ are left. With this algorithm, the large deviation function can be evaluated after each branching step as the deviation of the normalization,
\begin{equation}
\psi^t(\lambda) =  \ln \frac{1}{\Nw} \sum_{i=1}^{\Nw} W_i(t,n\delta t),
\end{equation}
which is an exponential average over the bias factors of each walker. In the limit of a large number of walkers, this estimate is unbiased \cite{hidalgo2017finite}.
The local estimate can be improved by averaging over the observation time,
\begin{equation}
\label{Eq:DMCpsi2}
\psi(\lambda) = \frac{1}{t_N} \sum_{t=1}^{t_N/(n\delta t)} \psi^t(\lambda)
\end{equation}
which upon repeated cycles of integration and population dynamics yields a statistically converged estimate of $\psi(\lambda)$. Alternatively, $\psi(\lambda)$ can be computed from histogram reweighting using Eq.~\ref{Eq:Rew} from the distribution of $J$s generated from each walker.  In the preceding, all calculations are integrated with LAMMPS \cite{plimpton1995fast} and where specified, combined with a diffusion Monte Carlo code called the Cloning Algorithm for Nonequilibrium Stationary States (CANSS) \cite{ray2017exact}. A detailed description of convergence criteria for this algorithm can be found in Reference \cite{ray2017importance}. 

\section{Results and Discussion}
To illustrate the utility of our method, we have tested it in three model transport processes. In Table 1, we list all the transport coefficients considered in this section, along with their corresponding Green-Kubo relations and the dynamical variables whose large deviation function we compute. For all of the models studied, we generate trajectories by integrating a Langevin equation of motion. A Markovian, stochastic equation is needed for the calculation of the large deviation function using the method presented in the previous section. For the position of particle $i$, denoted $\mathbf{r}_i = \{x_i,y_i,z_i\}$, this equation has the form
\begin{equation}
\label{Eq:Lang}
m_i \ddot{\mathbf{r}}_i =- \nabla_{\mathbf{r}_i} U(\mathbf{r}^N)-m_i\gamma \dot{\mathbf{r}}_i + \mathbf{R}_i \, ,
\end{equation}
where the dots denote time derivatives, $U(\mathbf{r}^N)$ is the total intermolecular potential from all $N$ particles at position $\mathbf{r}^N$, $m_i$ is the particle's mass, $\gamma$ is the frictional coefficient,  and $\mathbf{R}_i$ is a random force. The statistics of the random force is determined by the fluctuation-dissipation theorem, which for each component is
\begin{equation}
\left \langle R_i(t) \right \rangle=0 \, \quad\quad \left \langle R_i(t)R_j(t') \right \rangle=m_i k_{\mathrm{B}}T\gamma \delta(t-t')\delta_{ij}
\end{equation}
where $k_{\mathrm{B}}T$ is Boltzmann's constant times temperature, $\delta(t)$ is Dirac's delta function and $\delta_{ij}$ is the Kronicker delta. For all our calculations, we have chosen $\gamma$ carefully so that the thermostat has little effect on the transport properties of the system and we are able to recover response coefficients consistent with calculations done using Newtonian trajectories, when possible.

\subsection{Validation of methodology: shear viscosity}
To illustrate our methodology, we first consider the evaluation of the shear viscosity, $\eta$, which is typically easy to compute with traditional methods. The phenomenological law that defines the shear viscosity is  Newton's law of viscosity, which relates the shear stress of a fluid to an imposed shear rate,
\begin{equation}
\label{Eq:viscosity}
\sigma_{xy} = \eta \frac{\partial v_x}{\partial y},
\end{equation}
where $\sigma_{xy}$ is the $xy$-component of the stress tensor, and $({\partial v_x}/{\partial y})$ is the gradient of the $x$ component of velocity in the $y$ direction. The relevant molecular current for this process is the momentum flux, which is equivalent to $\sigma_{xy}$. The stress tensor is computable as 
\begin{equation}
\sigma_{xy} = \frac{1}{V} \left ( \sum_{i}m_i v_{xi}v_{yi}+\sum_{i}x_i f_{yi} \right ),
\end{equation}
where $V$ is the constant volume of the system, and $v_{ki}$ and $f_{ki}$ are the velocity and force exerted on particle $i$ in the $k$ direction, respectively. Given this identification of the current, its associated thermodynamic force is $X=(V/ \kB T)({\partial v_x}/{\partial y})$. From Eqs.~\ref{Eq:JLX} and \ref{Eq:viscosity} we can identify the relation between the shear viscosity and $L$ as $\eta=L (V/\kB T)$. 

\begin{figure}
\centering
\includegraphics[width=8.5cm]{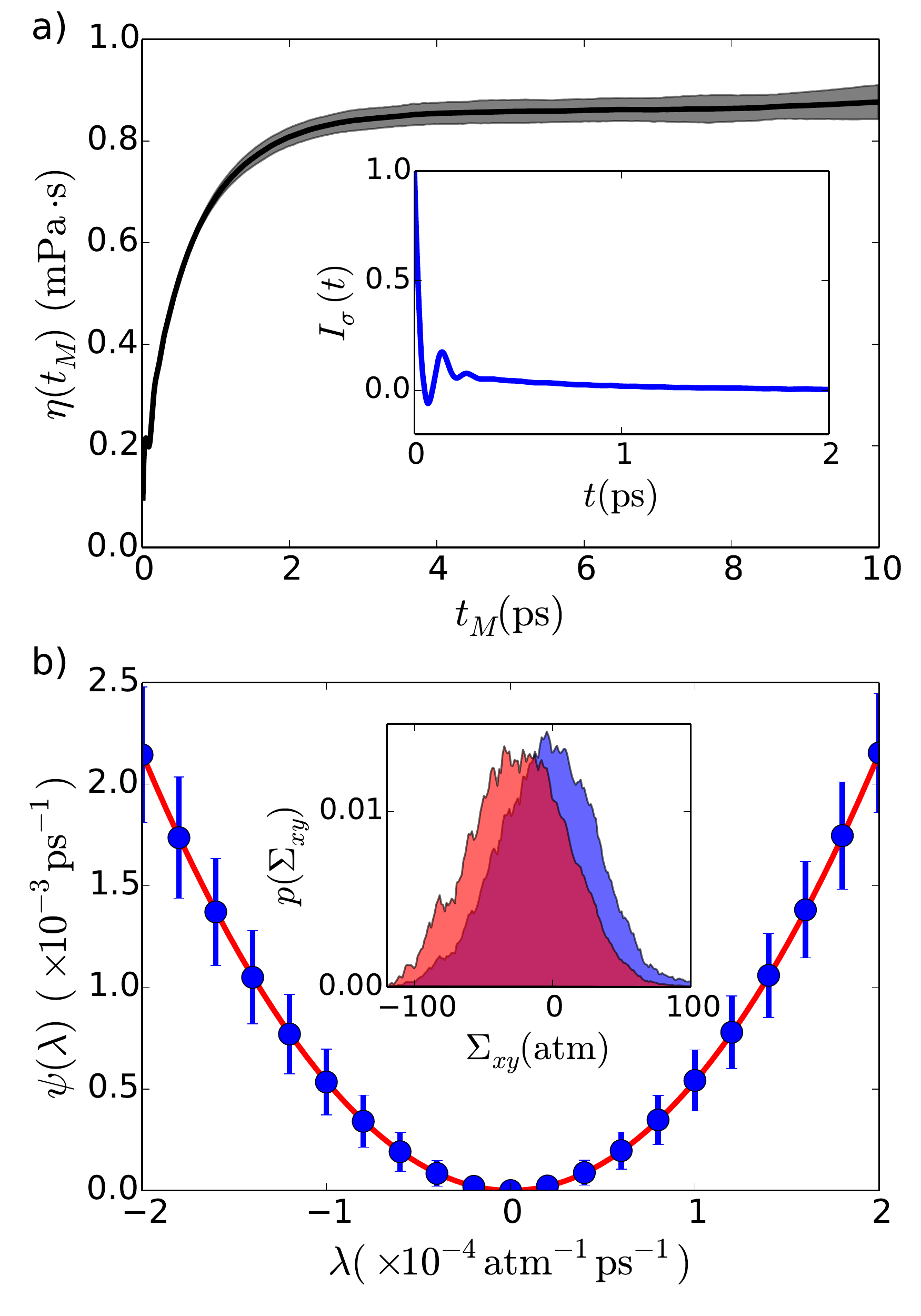}
\caption{(\textbf{a}) Shear viscosity of TIP4P/2005 water model as a function of the integration time $t_M$. Shading indicates the error bar computed from the standard error. The inset is the normalized autocorrelation function $I_\sigma(t)=\left \langle \sigma_{xy}(0)\sigma_{xy}(t) \right \rangle/\left \langle \sigma_{xy}(0)^2 \right \rangle$.  (\textbf{b})  Large deviation function for $\Sigma _{xy}$, as a function of the biasing parameter $\lambda$. The error bars indicate the standard error of the mean from 5 individual samples. The red line shows the parabolic fit of the data.  Inset is the original and the biased probability distribution of $\Sigma _{xy}$ at $\lambda=2\times 10^{-4} \textrm{atm}^{-1} \textrm{ps}^{-1}$.}
\label{Fi:Figure1}
\end{figure}   

We compute the shear viscosity for the TIP4P/2005 model of water \cite{abascal2005general}, which has been reported previously using Green-Kubo theory \cite{gonzalez2010shear}. Our simulation system consists of 216 water molecules with density $\rho$ = $1g/\mathrm{cm^3}$ and temperature $T$ = 298K, integrated with the Langevin equation in Eq.~\ref{Eq:Lang} with $\gamma=1$ $\textrm{ps}^{-1}$.  The simulation is thus done in an ensemble of constant number of molecules $N$, volume $V$, and temperature $T$. We have verified that for $\gamma=1$ $\textrm{ps}^{-1}$, the shear viscosity computed is the same as that from an ensemble with constant energy or an NVT ensemble using a Nos{\'e}-Hoover thermostat \cite{nose1984unified}. The molecules are held rigid with the SHAKE algorithm \cite{ryckaert1977numerical} and we employ a timestep of 1 fs. For all of the calculations, we first equilibrate the simulation for 20 ns. 

First, we compute $\eta$ using the Green-Kubo formula in Eq.~\ref{eta}. Note that other elements of the stress tensor can be averaged over to achieve better statistics. In both the Green-Kubo method and are new proposed calculation, the statistical benefit would be identical, so for notational clarity we will consider only the $xy$ component.
We average the stress-stress time correlation function over 20 ns, and this function is shown in Figure~\ref{Fi:Figure1}a. The time correlation function is oscillatory due to the inertial recoil of the dense fluid, and has largely decayed within 1 ps, though there is a slow component to the decay from the approximate conservation of momentum for times shorter than the timescale for the Langevin thermostat. From Green-Kubo theory, the viscosity is the integral of this function. Shown in the main part of Figure~\ref{Fi:Figure1}a, is $\eta(t_M)$ as a function of the upper limit of the integral as in Eq.~\ref{Eq:Tm}, which has plateaued by $t_M=10$ ps. Also shown are the associated statistical errors, which grow with $t_M$. The calculated shear viscosity from 5 independent simulations and a cutoff time of 20 ns is $0.876\pm 0.015$ $\textrm{mPa}\cdot \textrm{s}$. This value is in good agreement with that previously reported \cite{gonzalez2010shear}. 

Alternatively, we can compute the shear viscosity from the large deviation function for $\Sigma_{xy}$ defined in Eq.~\ref{Eta}. As the shear viscosity decays quickly for this model, importance sampling is unnecessary, so we illustrate the basic principle by brute force reweighting. Specifically, we generate an estimate of $p[\Sigma _{xy}]$ with $t_N= 80$ ps, from a 20 ns long equilibrium trajectory. Then, we reweight the distribution to compute  $p_\lambda [\Sigma _{xy}]$ according to Eq.~\ref{Eq:Rew}. Examples of the equilibrium and biased distributions are shown in the inset of Figure~\ref{Fi:Figure1}(b). The added bias shifts the distribution to a different mean, and the overlap between these two distributions determines the efficiency of our sampling. The large deviation function $\psi (\lambda)$, shown in the main panel in Figure~\ref{Fi:Figure1}(b), is evaluated by Eq.~\ref{Eq:LDF} for different  $\lambda$'s. The parabolic form of $\psi (\lambda)$ is in agreement with the Gaussian distribution of the fluctuation in $\Sigma _{xy}$ in the linear response regime. Given that  $\psi (\lambda)$ is a parabola centered at the origin, it is straightforward to compute $\eta$ from fitting the curve in Figure~\ref{Fi:Figure1}(b) to Eq.~\ref{Eq:EH} over a range of $|\lambda| \leq 1.5\times10^{-4}\textrm{atm}^{-1} \textrm{ps}^{-1}$. From this, we obtain an estimate of the viscosity $\eta =0.882\pm 0.017$ $\textrm{mPa}\cdot \textrm{s}$, which is in agreement with our Green-Kubo result. Both errors reported are the standard error of the mean.


\subsection{Analysis of systematic error: interfacial friction coefficient}
Having validated the basic methodology, we next focus on the systematic errors determining its convergence. As a case study, we consider computing the interfacial friction coefficient between a liquid-solid interface. This friction coefficient is defined by the linear relationship,
\begin{equation}
f_x=-\mu A v_s,
\end{equation}
where $f_x$ is the total lateral force exerted on the solid wall on the $x$ direction, $A$ is the lateral area of the interface, and $v_s$ is the tangential velocity of the fluid relative to the solid. As before, we can identify a relevant molecular current as the momentum flux along the wall, in this case proportional to
\begin{equation}
f_x = -  \sum_{i=1}^{N_l}\sum_{k=1}^{N_c} \frac{d}{dx_i} u_{ls}(|\mathbf{r}_i - \mathbf{r}_k|),
\end{equation} 
the sum of the $x$ component of the forces of all $N_l$ liquid particles on the $N_c$ wall particles, where the force is given by the gradient of the liquid-solid interaction potential, $u_{ls}$. Given this current, we can identify its conjugate force as $X=(A/ \kB T)v_s$, and consequently, the friction coefficient is given by $\mu = L (A/\kB T)$.

The system is modeled as a fluid of monatomic particles confined between two stationary atomistic walls parallel to the $xy$ plane. The fluid particles interact through a Lennard-Jones (LJ) potential with characteristic length scale $d$, energy scale $\epsilon$, time $\tau =\sqrt{md^2/\epsilon}$ with $m$ as the mass of the fluid particle, and is truncated at 2.5$d$. Reduced units will be used throughout this and the following section, and we set $k_{\mathrm{B}}=1$. The walls are separated by a distance $H_z=18.17d$ along the $z$ axis. Periodic boundary conditions are imposed along $x$ and $y$ directions, with the lateral dimensions of the simulation domain $H_x=H_y=15.90d$. Each wall is constructed with 1568 atoms distributed as (111) planes of face-centered-cubic lattice with density $\rho _w=2.73d^{-3}$, while the fluid density is $\rho _f=0.786d^{-3}$. The wall atoms do not interact with each other, but are allowed to oscillate about their equilibrium lattice sites under the harmonic potential $u_h(r)=k r^2/2$, with a spring constant $k=600\epsilon /d^2$. The mass of the wall atoms is chosen to be $m_c=4m$. The interaction between the wall and the fluid atoms is also modeled by a LJ potential with the same length scale $d$ and truncation, but a slightly smaller energy $\epsilon _{wf}=0.9\epsilon$, to model the solvophobicity of the wall \cite{sendner2009interfacial}. Only the wall particles are thermostatted by the Langevin equations in Eq.~\ref{Eq:Lang} using $\gamma=1\tau^{-1}$.

\begin{figure}
\centering
\includegraphics[width=8.5cm]{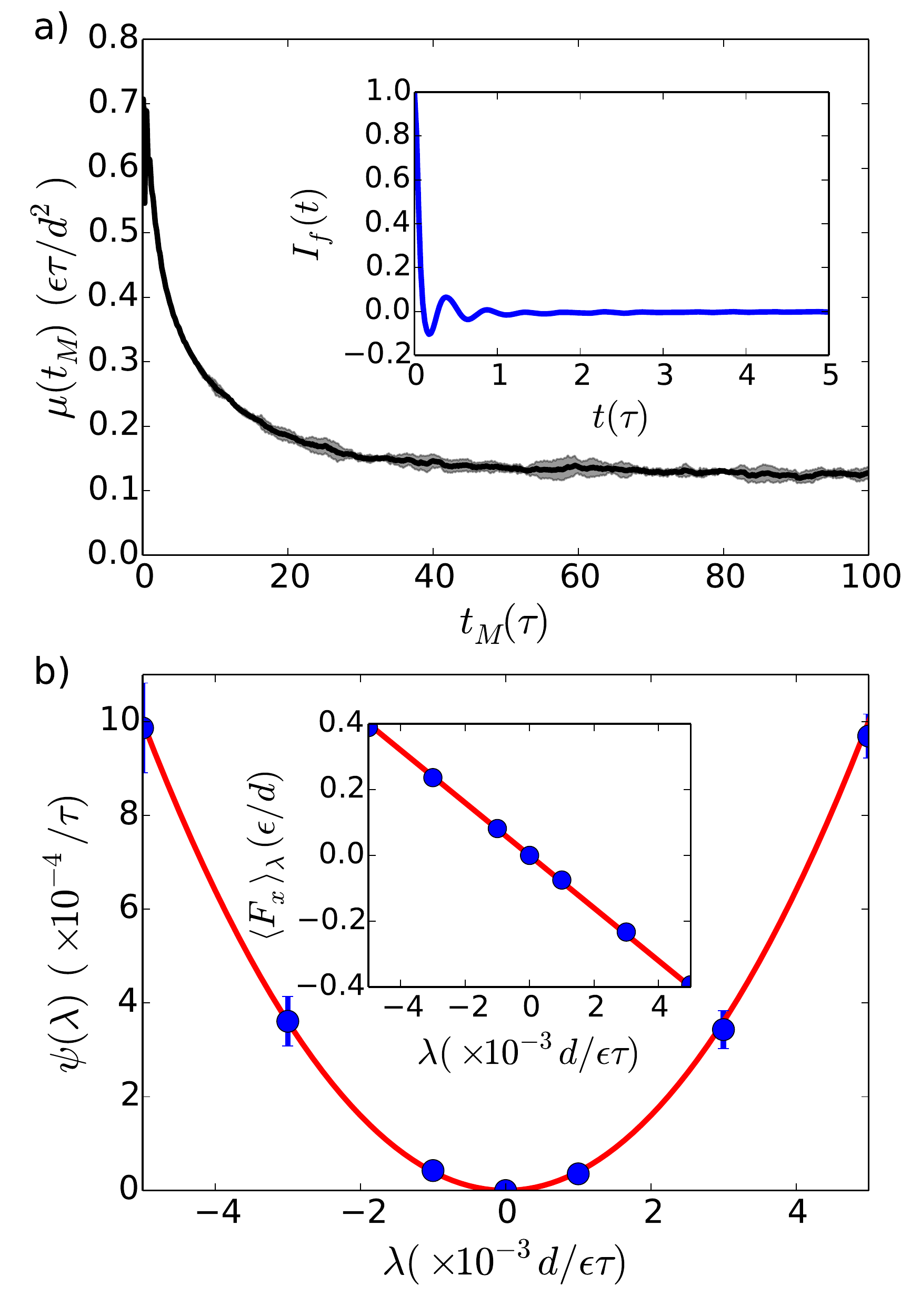}
\caption{ (\textbf{a}) Interfacial friction coefficient computed from Green-Kubo method as a function of the integration time $t_M$. Inset is the normalized force autocorrelation function $I_f(t)=\left \langle f_{x}(0)f_{x}(t) \right \rangle/\left \langle f_{x}(0)^2 \right \rangle$. (\textbf{b}) Large deviation function of the dynamical observable $F_x$ with $t_N=400\tau$. Red line is the parabolic fit. The inset is the average observable $\left \langle F_x \right \rangle _{\lambda}$ in the biased ensemble, with the linear fit in red.}
\label{Fi:Figure2}
\end{figure}   

Previous studies have recognized that $\mu$ is difficult to compute due to the confinement of the corresponding hydrodynamic fluctuations \cite{petravic2007equilibrium,huang2014green}, which results in a large systematic error. This difficulty has led to some questioning the reliability and applicability of Green-Kubo calculations, such as the one derived in \cite{bocquet2013green} and shown in Eq.~\ref{mu}, to compute $\mu$. Indeed, we have found that the details of the simulation, such as the ensemble, system geometry and $\gamma$ used in the Langevin thermostat, all have an important influence on the calculation of $\mu$. This sensitivity is because the fluctuations that determine the friction are largely confined to two spatial dimensions, which is well known to result in correlations that have hydrodynamic long time tails, whose integral may be divergent \cite{alder1970decay}. However, both our large deviation function method and the Green-Kubo calculations are based on equilibrium fluctuations. Provided a simulation geometry, equation of motion, and ensemble, the system samples the exact same trajectories, so we expect the friction coefficient computed in both ways to agree.
 Shown in the inset of Figure~\ref{Fi:Figure2}(a) is the Green-Kubo correlation function, which includes a very slow decay extending to at least 100 $\tau$, following short time oscillatory behavior from the layered density near the liquid-solid interface. 
The main panel of Figure~\ref{Fi:Figure2}(a) shows $\mu$ computed with increasing integration time, $t_M$. 
Averaging over 4 independent samples with a cutoff $t_M=1000\tau$, our estimation of the friction coefficient is $\mu=0.109\pm0.019$ $\epsilon\tau/d^2$. 
%
%
The interfacial friction coefficient is also computed from the large deviation function, with $t_N=400\tau$, using the time integrated force, Eq.~\ref{Mu}, as our dynamical observable.
The large deviation function and the average time integrated force, $\left \langle F_x \right \rangle_\lambda$, are shown in the main panel and inset of Figure~\ref{Fi:Figure2}(b), respectively, demonstrating that within the range of $\lambda$ we consider the system still responds linearly. With $\lambda=10^{-3}\sigma/\epsilon\tau$ and $t_N=4000\tau$,
importance sampling gives us an estimate of the friction coefficient as $\mu=0.121\pm0.002$ $\epsilon\tau/d^2$, in reasonable agreement with the Green-Kubo estimate and with previous reports~\cite{petravic2007equilibrium}.

\begin{figure}
\centering
\includegraphics[width=8.5cm]{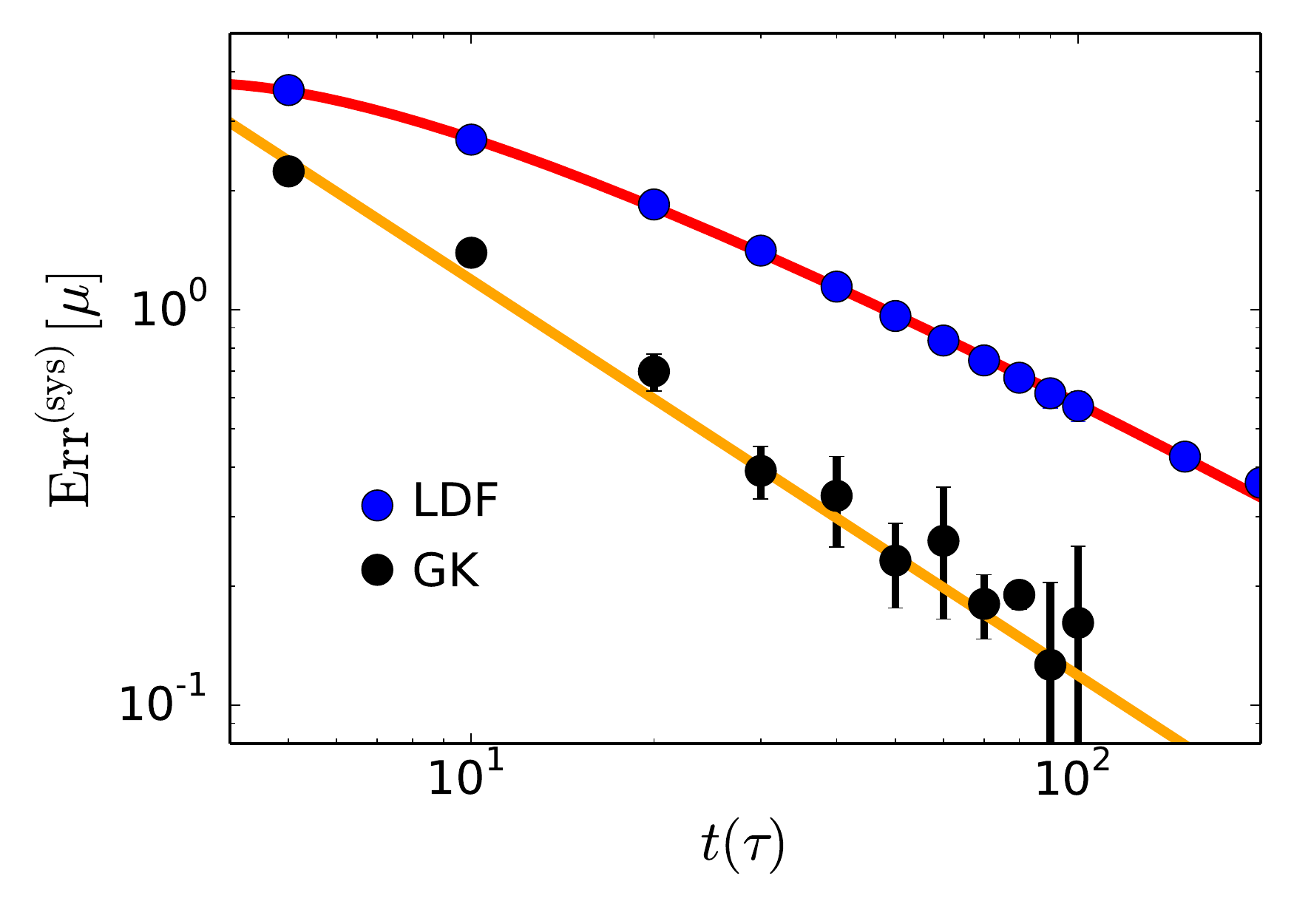}
\caption{ Relative systematic error $\mathrm{Err}^{(\textrm{sys})}[\mu]$ due to finite time in the estimation of $\mu$ in the Green-Kubo (GK) method (black) and the large deviation function (LDF) method (blue). The time, $t$, in the $x$-axis denotes the upper limit of the integral, $t_M$, in the Green-Kubo method, or the total length of the trajectory, $t_N$, in computing the large deviation function. The red line is a fit to the function $y=a \ln(b t)/t$, and the orange line is a fit to $y=a/t$. }
\label{Fi:Figure3}
\end{figure}

In both the Green-Kubo and the large deviation function calculations, the main source of systematic error is from finite time. This error is especially highlighted in this example, where the time correlation function decays very slowly. We consider the systematic errors in the estimate of $\mu$ by defining a relative error as
$$
\mathrm{Err}^{(\textrm{sys})}[\mu]=(\mu(t)-\mu )/\mu,
$$
where $\mu(t)$ is the finite time value of the friction coefficient, and $\mu$ its asymptotic value at $t\rightarrow\infty$. The form of the time dependent systematic error is different in the Green-Kubo method compared to the large deviation estimate. In the Green-Kubo method, systematic errors come from truncating the integral before the correlation function has decayed, and we denote this time $t_M$, the cutoff time in the integral of the correlation function. In the large deviation calculation, systematic errors come from both truncating the integral as well as sub-time-extensive contributions to the exponential expectation value, which are more analogous to finite size effects in normal free energy calculations. These contributions are both determined by the path length $t_N$. The relative systematic error is shown in Figure~\ref{Fi:Figure3} for both methods. For this case, it appears that the Green-Kubo method always has a smaller error than the large deviation function method, though their magnitudes are comparable. 

In the Green-Kubo method, it follows that if we know the analytical form of the correlation function, we can determine the scaling of the relative error. In the case of interfacial friction, Barrat and Boquet have proposed that for a cylindrical geometry where the dimension on the confined direction is much smaller than the other two direction, the force autocorrelation should decay asymptotically as $\sim 1/t^2$ using hydrodynamic arguments \cite{bocquet2013green}. This is a direct consequence of the fact that the velocity autocorrelation function decays as $\sim 1/t$ in a 2-dimensional system \cite{alder1970decay}, neglecting the self-consistent mode coupling correction that adds an imperceptible $\sqrt{\ln t}$ correction \cite{wainwright1971decay,isobe2008long}. This is confirmed in our simulation result in Figure~\ref{Fi:Figure3} (orange line), where the integral of the force correlation function decays as $\sim 1/t$.
 
Since the large deviation function has a Gaussian form, we can analyze the form of the finite time correction exactly as
\begin{eqnarray}
\label{Eq:SysErr}
\mathrm{Err}^{(\textrm{sys})}[\psi] &=&\frac{\tilde{\psi}(\lambda,t_N)-\psi(\lambda)}{\psi(\lambda)} \\
&=&\frac{\mu(t_N)-\mu}{\mu}+\frac{1}{2t_N\mu\lambda^2}\ln[4\pi t_N\mu(t_N)], \nonumber
\end{eqnarray}
where $\psi(\lambda)$ is the long time limit of the large deviation function, and $\tilde{\psi}(\lambda,t_N)$ is its finite time estimate. This follows from a fluctuation correction about a saddle point integration. Physically, this correction arises from a $t_N$ that is too short, such that $\psi(\lambda)$ is not the dominant contribution to the tilted propagator, but rather includes temporal boundary terms from the overlap of the distribution of initial conditions and the steady state distribution generated under finite $\lambda$ \cite{hidalgo2017finite}. If we expand the first term, we arrive at
\begin{equation}
\mu(t_N)-\mu\approx-\int_{t_N}^{\infty}\left \langle j(0)j(t) \right \rangle dt+\frac{1}{t_N}\int_0^{t_N}t\left \langle j(0)j(t) \right \rangle dt,
\end{equation} 
which consists of the term included in the Green-Kubo expression, as well as an additional term modulated by a factor of $1/t_N$. Given that the correlation decays as $\sim 1/t^2$, the first term on the right hand side scales as $\sim 1/t_N$, as in the Green-Kubo method, while the second term scales as $\sim (1/t_N\ln t_N)$. This form is shown in Figure~\ref{Fi:Figure3} and agrees very well with our data. These additional terms explain why the magnitude of the systematic error is larger for the large deviation function. In cases where the Green-Kubo correlation function decays faster than $1/t^2$, we expect that the dominant contribution to the error will come from the last term in Eq.~\ref{Eq:SysErr}.  

\subsection{Analysis of statistical error:\\thermal conductivity}

We finally discuss the statistical error of our method by studying the thermal conductivity, $\kappa$, of a solid system with particles that interact via the Weeks-Chandler-Anderson potential \cite{weeks1971role}. The thermal conductivity is defined through Fourier's law,
\begin{equation}
\mathbf{e}=-\kappa \mathbf{\nabla T},
\end{equation}
where $\mathbf{e}$ is the energy current per unit area, and $\mathbf{\nabla T}$ is the temperature gradient. From the expression for entropy production, the thermodynamic force is given by $X=-(1/k_{\mathrm{B}}T^2)\nabla T$, and so the thermal conductivity $\kappa=L/(Vk_{\mathrm{B}}T^2)$. As the relevant molecular current, we study the fluctuations of the heat flux $\mathbf{q}$ given by
\begin{equation}
\mathbf{q}=\mathbf{e}V=\sum_{i}\mathbf{v_i}e_i+\frac{1}{2}\sum_{i\neq k}(\mathbf{f_{ik}}\cdot \mathbf{v_i})\mathbf{r_{ik}},
\end{equation}
where $e_i$ is the per-particle energy, $\mathbf{f_{ik}}$ is the force on atom $i$ due to its neighbor $k$ from the pair potential, and $\mathbf{r_{ik}}$ is the coordinate vector between the two particles. We use a system size of $10^3$ unit cells, with lattice spacing 1.49$d$. A Langevin thermostat with $\gamma =0.01\tau ^{-1}$ maintains the system at the state point $T =1.0\epsilon /k_{\mathrm{B}}$, $\rho =1.2d^{-3}$, which yields identical results for $\kappa$ as an NVE calculation. We focus on the diagonal component, $\kappa_{xx}$, of the thermal conductivity tensor.

\begin{figure}
\centering
\includegraphics[width=8.5cm]{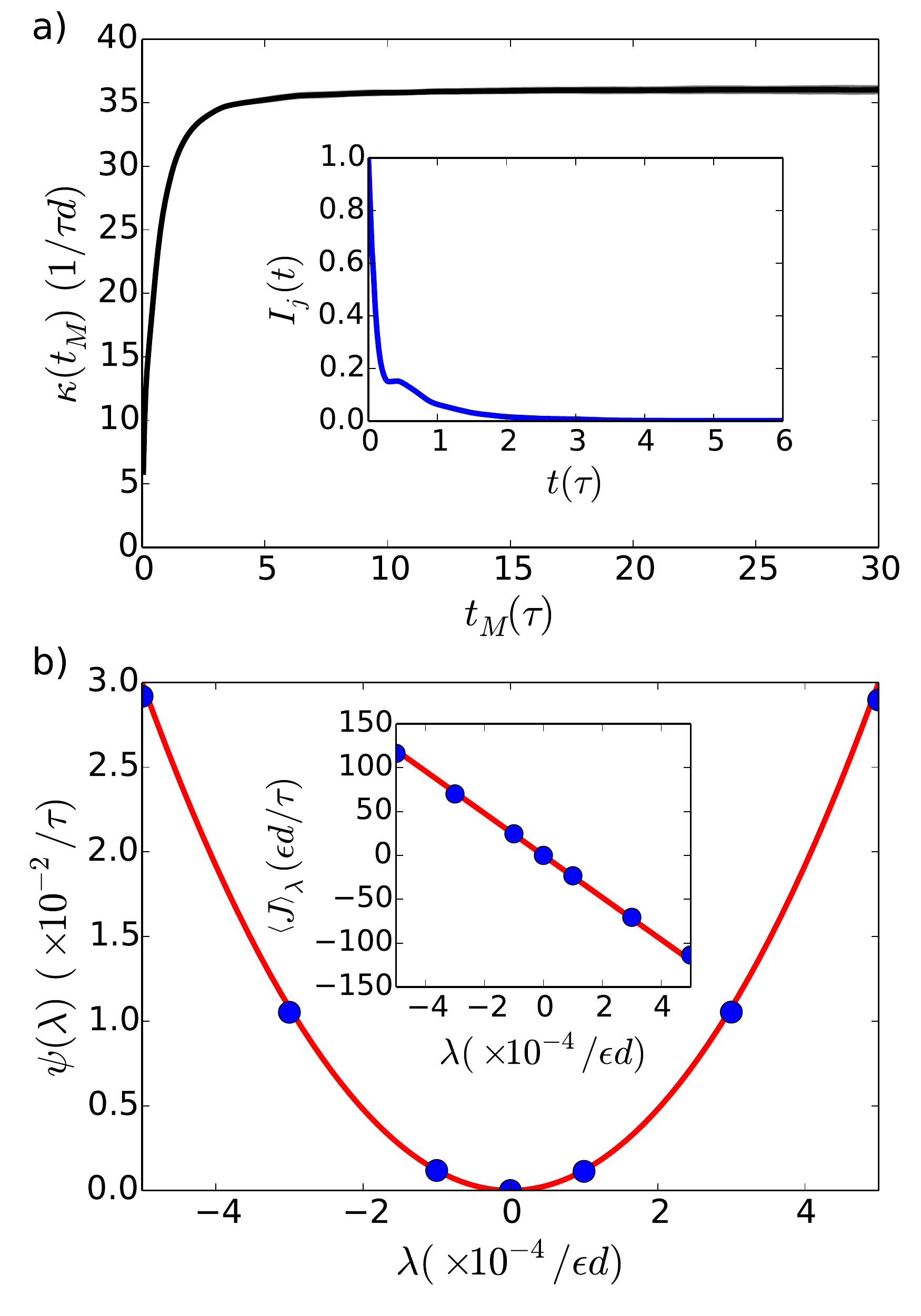}
\caption{ Calculation of the thermal conductivity, $\kappa$, of a WCA solid at $T=1.0\epsilon/k_{\mathrm{B}}$, $\rho=1.2d^{-3}$. (\textbf{a}) $\kappa(t_M)$ calculated by integrating the heat flux correlation function up to time $t_M$. The data are averaged from 4 samples and the error bars are standard deviations, which are smaller than the symbols. The inset shows the normalized heat flux correlation function, $I_q(t)=\left \langle q_{x}(0)q_{x}(t) \right \rangle/\left \langle q_{x}(0)^2 \right \rangle$. (\textbf{b}) Large deviation function of dynamical observable, $Q_x$, as a function of the bias $\lambda$. The red line is the parabolic fit. The inset is the average observable in the biased ensemble, $\left \langle Q \right \rangle _{\lambda}$, as a function of $\lambda$ with the linear fit in red.
 }
\label{Fi:Figure4}
\end{figure}

Within Green-Kubo theory, the thermal conductivity can be computed by integrating the autocorrelation function of the $x$ component of the heat flux, $q_x$, as in Eq.~\ref{kappa}. The inset of Figure~\ref{Fi:Figure4}(a) is the decay of the autocorrelation function, which comprises a fast decay from the high-frequency vibrational modes, followed by a slower decay that contributes most to the thermal conductivity and arises due to the low frequency acoustic modes \cite{che2000thermal}. To compute $\kappa$ from the integral, as shown in the main part of Figure~\ref{Fi:Figure4}(a), the upper time limit is chosen as $t_M=1500\tau$, though the relaxation of the correlation extends only to around 5$\tau$. To compute $\kappa$ from the large deviation function, we study fluctuations in the time averaged heat flux, $Q_x$, defined in Eq. ~\ref{Kappa}. The transport coefficient, $\kappa$, is again calculated using Eq.~\ref{Eq:EH} by assuming the large deviation function $\psi (\lambda)$ as a parabola, which is justified in Figure~\ref{Fi:Figure4}(b). The inset there shows clearly the linear response of the biased ensemble average, $\left \langle Q \right \rangle _{\lambda}$, computed from Eq.~\ref{Eq:BiasedEn}. Given sufficient statistics the two methods converge to the same value. The estimate of thermal conductivity from the Green-Kubo method using a long trajectory of $1.5\times10^6$ $\tau$ is $\kappa=34.3\pm2.2$ $1/\tau d$, while the estimate from the large deviation function using $\Nw=1000$ walkers and $\lambda=10^{-4}$ is $\kappa=34.01\pm0.78$ $1/\tau d$. 

\begin{figure}
\centering
\includegraphics[width=8.3cm]{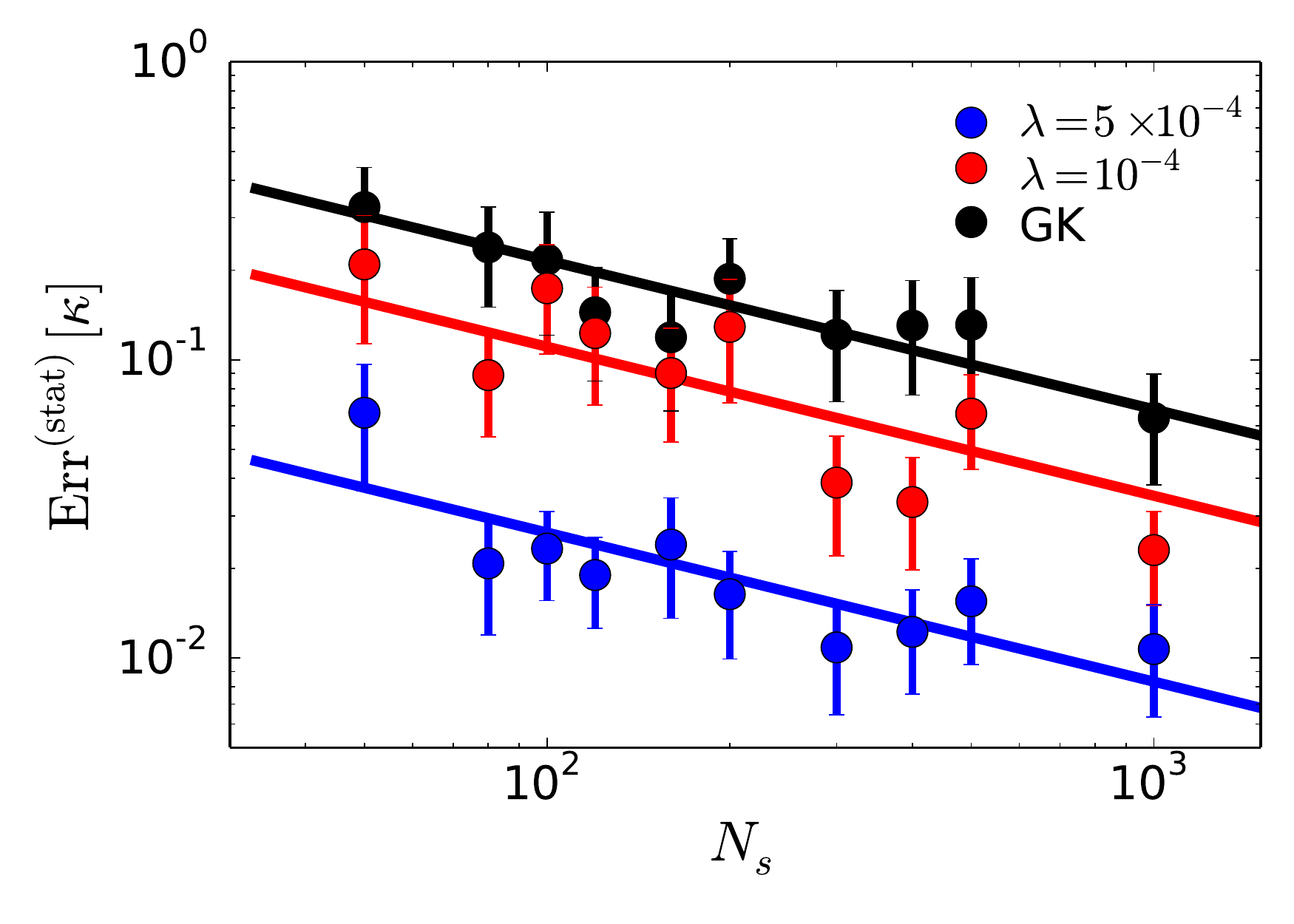}
\caption{ Relative statistical error in the measurement of $\kappa$, the Green-Kubo (GK) method (black) and the large deviation function method with $\lambda=10^{-4}$ (red) and $\lambda=5\times 10^{-4}$ (blue).  $N_s$ denotes the number of walkers $N_w$ used in evaluating the large deviation function, or $N_a$, an indicator of the total averaging time in the Green-Kubo method. The solid lines are fits of function $y=a/\sqrt{N_s}$.}
\label{Fi:Figure5}
\end{figure}

While the average values of $\kappa$ agree between the two methods, the statistical convergence varies significantly. To make a fair comparison, we set the total observation time of the trajectories to the same time as the upper limit of the Green-Kubo integral, i.e. $t_N=t_M=1500\tau$, which is much longer than the characteristic decay of the current autocorrelation function. To compensate for computational overhead of propagating $\Nw$ trajectories in parallel in the cloning algorithm, the total averaging time of the Green-Kubo method is chosen as $t_{\mathrm{tot}}=t_M\times N_a$, and $N_a$ equals the walker number, $N_w$, so that the two methods require approximately the same computational effort. Both $N_a$ and $N_w$ will be denoted as $N_s$ reflecting the number of independent samples of each fluctuating quantity. We measure the statistical error by the relative error
\begin{equation}
\mathrm{Err}^{(\textrm{stat})}[\kappa]=\frac{\sqrt{\left \langle \delta\kappa^2 \right \rangle}}{\kappa},
\end{equation}
which is plotted in Figure~\ref{Fi:Figure5} for both methods. As usual, the statistical error depends on both the relative size of observable fluctuations, and the number of independent samples. We find that as the standard deviations of both methods scale as $1/\sqrt{N_s}$ as expected, our importance sampling clearly helps to suppress the statistical error compared to the Green-Kubo method with similar computational effort, decreasing the magnitude of the error by an order of magnitude at fixed $N_s$. Even though we have to choose a bias small enough to guarantee a linear response, we do see that larger bias helps to yield statistically reliable results. 

Jones and Mandadapu have performed a rigorous error analysis on the estimates of Green-Kubo transport coefficients with the assumption that the current fluctuations follow a Gaussian process \cite{jones2012adaptive}. They found that the variance of $\kappa$ is a monotonically increasing function of $t_M$, and arrived at an upper bound for the relative error
\begin{equation}
\mathrm{Err}^{(\textrm{stat})}[\kappa]<2\sqrt{\frac{t_M}{t_{\mathrm{tot}}}}=2\sqrt{\frac{1}{N_a}} \,
\end{equation}
which depends on only the number of trajectory segments of length $t_M$. As a consequence, the statistics become worse when the system has longer correlation times, and there are no ways of controlling the intrinsic variance of the observable. On the other hand, in the large deviation method, the relative error in the large deviation function is
\begin{eqnarray}
\mathrm{Err}^{(\textrm{stat})}[\psi(\lambda)] &=&\frac{1}{\psi(\lambda)}\sqrt{\frac{\psi^{''}(\lambda)}{N_w}} \\
&=&\frac{1}{\lambda^2}\sqrt{\frac{2}{L N_w}} \quad\quad |\lambda| > 0 \nonumber
\end{eqnarray}
which depends on not only the number of samples, in this case $\Nw$, but also has a dependence on $\lambda$ and $L$. In general, as $\lambda$ increases, the walkers will become more correlated. However, within the regime of linear response, or to first order in $\lambda$, the number of uncorrelated walkers should be $N_w$. Because the large deviation function, $\psi(\lambda)$, scales as $\lambda ^2$ while its second derivative, $\psi^{''}(\lambda)$, has no dependence on $\lambda$, the relative size of the fluctuations can be tuned by changing  $\lambda$ away from 0. This is verified in Figure~\ref{Fi:Figure5}, where increased $\lambda$ generates an order of magnitude reduction in the statistical error relative to the Green-Kubo calculation. This decrease in the statistical error is also confirmed for a series of $\lambda$'s. This tunability afforded by the large deviation function calculation is the same advantage afforded by direct simulation of transport processes where the relative size of fluctuations is determined by the size of the average current produced by driving the system away from equilibrium. Instead of evaluating $\kappa$ from the large deviation function directly, we could have derived it from the change in the average current produced at a given $\lambda$. However, in such a case, the relative error would only scale as $|\lambda|$ rather than $\lambda^2$.

\section{Conclusions}
In this paper, we have explored the possibility of calculating transport coefficients from a large deviation function or a path ensemble free energy. The robustness of our method is tested by a variety of model systems ranging in composition and complexity of molecular interactions. 
Our method is general, and we expect the addition of importance sampling to be beneficial in instances where statistical errors are dominant. More precisely, our analysis shows that the systematic errors for both the Green-Kubo calculation and the large deviation calculation are asymptotically the same if the time correlation function decays faster than $1/t^2$. If the correlation function decays slower, than there will be a larger systematic error for the large deviation function calculation that will need to be converged at large $t_N$. In such cases, the form of this error follows from Eq.\ref{Eq:SysErr} and scales as $1/t_N \ln t_N$. Such slow decay is expected for low-dimensional systems where the current includes hydrodynamic modes. Our analysis of the relative statistical errors between the Green-Kubo and the large deviation function calculations show that our method requires generically fewer statistically uncorrelated samples for comparable statistical accuracy.  This is a consequence of the importance sampling employed. The magnitude of this statistical efficiency, defined as the number of independent samples needed for a given error ($N_w/N_a$) increases linearly with the size of the transport coefficient, $L$ and increases rapidly with the increasing bias, as $\lambda^4$.
%

While we have considered only linear response coefficients, our method can be easily extended to the nonlinear regime or to off-diagonal entries in the Onsager matrix, where Green-Kubo formulas are even more cumbersome to evaluate and few direct methods exist or can be formulated. These extensions are possible since the diffusion Monte Carlo algorithm is capable of sampling rare fluctuations in the non-Gaussian tails of the distribution. Moreover, it is also possible to probe the response around nonequilibrium steady states, as the method presented here does not rely on an underlying Boltzmann distribution. 
\vspace{6pt} 


\acknowledgments{D.T.L. and C.Y.G. was supported by the UC Berkeley College of Chemistry. The authors would like to thank Ushnish Ray for useful discussions and for developing the use of LAMMPS with the CANSS package, available at https://github.com/ushnishray/CANSS, for the calculation of nonequilibrium properties of complex systems.}


\end{document}